\documentclass[10pt,conference]{IEEEtran}
\IEEEoverridecommandlockouts
\usepackage{cite}
\usepackage{amsmath,amssymb,amsfonts}
\usepackage{graphicx}
\usepackage{textcomp}
\usepackage{xcolor}

\usepackage{color}

\usepackage{soul}
\usepackage{xspace}

\newcommand{\noindentparagraph}[1]{\par\noindent\textbf{#1}.}

\usepackage{comment}

\usepackage{makecell}

\usepackage{cite}
\usepackage{amsmath,amssymb,amsfonts}
\usepackage{algorithm}
\usepackage{algpseudocode}

\usepackage{graphicx}
\usepackage{textcomp}
\usepackage{xcolor}
\usepackage{hyperref} 
\def\BibTeX{{\rm B\kern-.05em{\sc i\kern-.025em b}\kern-.08em
    T\kern-.1667em\lower.7ex\hbox{E}\kern-.125emX}}
\usepackage{booktabs} 
\usepackage{url}
\usepackage{xspace}
\usepackage{caption}
\usepackage{subcaption}
\usepackage{float}
\usepackage{textcomp}
\usepackage{multirow,makecell}
\usepackage{pifont}
\usepackage{tikz}
\usepackage{bbm}

\usepackage{paralist}

\newcommand{\etal}{{\em et al.}\xspace}
\newcommand{\ie}{{\em i.e.,}\xspace}
\newcommand{\eg}{{\em e.g.,}\xspace}

\usepackage{amsmath}
\usepackage{graphicx}
\usepackage{makecell}

\usepackage{colortbl}
\newcommand{\cc}{\cellcolor{black!5!white}}
\newcommand{\ccd}{\cellcolor{blue!5!white}}
\usepackage{hhline}

\usepackage{framed}
\usepackage{array}
\newcolumntype{H}{>{\setbox0=\hbox\bgroup}c<{\egroup}@{}}

\def\BibTeX{{\rm B\kern-.05em{\sc i\kern-.025em b}\kern-.08em
    T\kern-.1667em\lower.7ex\hbox{E}\kern-.125emX}}

\renewcommand{\paragraph}[1]{\par\textit{#1}:}

\begin{document}

\title{I Can Find You in Seconds! Leveraging Large Language Models for Code Authorship Attribution}

\author{
  \IEEEauthorblockN{Soohyeon Choi\IEEEauthorrefmark{1}\IEEEauthorrefmark{2},
    Yong Kiam Tan\IEEEauthorrefmark{2}\IEEEauthorrefmark{3}, 
    Mark Huasong Meng\IEEEauthorrefmark{4},
    Mohamed Ragab\IEEEauthorrefmark{5},\\ 
    Soumik Mondal\IEEEauthorrefmark{2},
    David Mohaisen\IEEEauthorrefmark{1} and 
    Khin Mi Mi Aung\IEEEauthorrefmark{2}
  }
\IEEEauthorblockA{\IEEEauthorrefmark{1}University of Central Florida, USA}
\IEEEauthorblockA{\IEEEauthorrefmark{2}Institute for Inforcomm Research (I$^2$R), A*STAR, Singapore}
\IEEEauthorblockA{\IEEEauthorrefmark{3}Nanyang Technological University, Singapore}
\IEEEauthorblockA{\IEEEauthorrefmark{4}Technical University of Munich, Germany}
\IEEEauthorblockA{\IEEEauthorrefmark{5}Propulsion and Space Research Center, Technology Innovation Institute, UAE}
}

\maketitle

\begin{abstract}
Source code authorship attribution has received attention from the security and software engineering research communities due to its potential uses in software forensics, plagiarism detection, and protection of software patch integrity.
Existing code authorship attribution techniques mainly resort to supervised machine learning techniques, which rely heavily on extensive labeled datasets yet struggle with generalization across diverse programming languages and coding styles.
This paper is inspired by recent advancements in natural language authorship analysis brought about by large language models (LLMs)---LLMs have demonstrated remarkable performance and generalization in various tasks without task-specific tuning or pre-training on labeled data. Thus, we seek to similarly leverage LLMs to address the technical challenges of source code authorship tasks.  

We design and present a comprehensive empirical study showing that state-of-the-art LLMs are capable of attributing source code authorship across different programming languages.
Specifically, they offer promising performance in determining whether two source codes are written by the same individual with zero-shot prompting, achieving Matthews Correlation Coefficient (MCC) score up to 0.78; they can also attribute code authorship from a small group of reference code snippets through few-shot in-context learning, achieving MCC up to 0.77; and, they also offer a degree of adversarial robustness against state-of-the-art misattribution attacks.
Despite these capabilities, we observed that na\"ive prompting of LLMs in code authorship attribution does not scale against the number of authors due to the LLMs' inherent input token limitations.
To circumvent this limitation, we propose a simple but effective tournament-style approach to leverage LLMs for code attribution over a large number of authors.
\textcolor{black}{We evaluate the approach on datasets written in C++ (500 authors, 26,355 code samples) and Java (686 authors, 55,267 code samples), crawled from Github as of November 2024. The results show that the proposed approach can accurately attribute code authorship even in real-world, few-shot settings achieving classification accuracy of up to 65\% for C++ and 68.7\% for Java using only one reference code per author.}
These findings open new avenues for leveraging LLMs in code authorship attribution tasks with applications in cybersecurity and software engineering.

\end{abstract}


\section{Introduction}\label{sec:introduction}
Source code authorship attribution is the task of determining the author(s) of a piece of source code written in a specific programming language \cite{kalgutkar2019code,burrows2007source,choi2023untargeted, choi2024chatgpt}.
It has received wide attention for its potential use cases in software engineering and security. For instance, it can be used to safeguard software intellectual property against copyright infringement~\cite{alsulami2017source} and ensure the integrity and authenticity of code modifications throughout the software life cycle~\cite{bogomolov2021authorship}.
From the security perspective, source code authorship attribution can be used to trace and find the programmer(s) of malicious code, thereby assisting in software forensics for cybercrime investigations and also prevention of further threats~\cite{alrabaee2017feasibility,kalgutkar2019code}.

Classical methods for code authorship attribution mainly rely on machine learning (ML) and deep learning (DL) techniques in order to analyze linguistic and structural characteristics of the source code~\cite{abuhamad2018large, caliskan2015anonymizing,abuhamad2020multi,bogomolov2021authorship}.
However, the use of supervised ML/DL methods necessitates a costly and time-consuming training process involving a vast amount of author-labeled training data to attain acceptable accuracy.
The trained models' performance is also closely tied to the quality and coverage of the training data, \eg the distribution of authors and programming languages available in the data~\cite{abuhamad2018large,caliskan2015anonymizing}. These challenges limit the generalization ability of pre-trained ML/DL models, especially on unseen authors and languages, where they cannot produce any classification result.

Recently, large language models (LLMs) such as Gemini~\cite{gemini}, ChatGPT~\cite{ChatGPT}, and Llama~\cite{llama} have gained widespread popularity as foundation models in many applications because of their user-friendly conversational interface and impressive human-like language capabilities~\cite{zhao2023survey}.
These LLMs benefit from large model sizes and extensive training processes, and they have demonstrated remarkable performance in handling diverse domain-specific tasks such as natural language translation~\cite{xie2023translating}, creative writing~\cite{yuan2022wordcraft}, and source code comprehension and generation~\cite{xu2022systematic,leinonen2023using}.

In this work, we are inspired by the recent advancement in natural language authorship analysis brought about by LLMs~\cite{huang2024can}, and we seek to leverage LLMs to address the existing technical challenges of source code authorship attribution tasks.
To this end, we present the first empirical study covering four mainstream LLMs families namely ChatGPT~\cite{ChatGPT}, Gemini~\cite{gemini}, Mistral~\cite{jiang2023mistral7b}, and Llama~\cite{llama}, to explore whether they are capable of source code authorship attribution tasks given little or even no author-labeled references; these are considered to be particularly challenging settings for traditional ML/DL methods.
Specifically, we investigate the various LLMs' capacity for determining whether two code samples are written by the same author using a \emph{zero-shot} query  (\textbf{RQ1}) and for classifying the authorship of a code sample based on a small set of author-labeled code as reference through \emph{few-shot} in-context learning (\textbf{RQ2}).
The results of our empirical study demonstrate promising capabilities in state-of-the-art LLMs for code authorship attribution tasks.

Throughout the empirical study, we also observe that LLM-based authorship attribution by na\"ive prompting techniques does not scale against the number of candidate authors due to the inherent input token limitations of LLMs. 
To circumvent this challenge, we propose a simple but effective tournament prompting approach to perform attribution analysis across multiple rounds (\textbf{RQ3}).
\textcolor{black}{To avoid potential dataset leakage in the LLMs' training data (which may unintentionally boost our results), and to test our approach against a large number of authors on real-world data, we constructed datasets by crawling public GitHub repositories dated between May and October 2024, and ran our experiments in November 2024.}
\textcolor{black}{Our evaluation in this setting shows that the proposed tournament-style authorship approach can accurately attribute code authorship on a large scale with few-shot prompting---the approach reaches a Top-1 accuracy of 65\% when classifying over 500 C++ author candidates, using only a single reference code sample per author.}

Another important application of authorship attribution is in code forensics, where an effective attribution solution must account for adversarial settings.
However, existing ML/DL-based approaches have been shown to be vulnerable to misattribution attacks~\cite{quiring2019misleading,li2022ropgen}.
Therefore, we assess the robustness of LLMs against state-of-the-art adversarial misattribution attacks (\textbf{RQ4}).
Here, we find that the tested LLMs offer a promising level of robustness without the need to tailor or fine-tune our prompts. This robustness can be further enhanced using adversarial-aware prompting.

Finally, whereas ML/DL approaches need to be re-trained for each new programming language~\cite{caliskan2015anonymizing,abuhamad2018large}, we carry out additional evaluations to explore whether LLMs can generalize their authorship attribution capabilities to different languages (\textbf{RQ5}).
Our results show that the tested LLMs can be readily applied to a different programming language (Java) with unchanged prompts; our proposed tournament prompting approach for large-scale problems is also language-agnostic. 

We summarize the contributions as follows:
\begin{itemize}
    \item In Section~\ref{sec:empirical}, we design and present an empirical study of code authorship attribution capabilities for state-of-the-art LLMs.
    Our study includes both zero-shot and few-shot prompting methods to perform various authorship attribution tasks.
    The results demonstrate LLMs' promising capabilities in code authorship tasks without reliance on extensive datasets and expensive training processes.
    \item In Section~\ref{sec:ours}, we propose a tournament-style approach to address the inherent input token limitations of LLMs. Our approach takes advantage of few-shot in-context learning to precisely attribute code authorship on a large scale.
    \item In Section~\ref{sec:evaluation}, we examine the robustness and generalization of LLM-based code authorship by, respectively, testing against state-of-the-art misattribution attacks and evaluating our approach on different programming languages.
    We demonstrate robustness and generalization capabilities without the need to tailor or tune our prompts.
\end{itemize} 

Beyond answering the above-mentioned research questions, we discuss further insights from the evaluation and provide potential future directions for improvements for our LLM-based approach. 
These findings open new avenues for leveraging LLMs in source code authorship attribution tasks with potential applications in broader areas of software engineering and cybersecurity.


\section{Background and Related Work}\label{sec:related}
\subsection{Code Authorship Attribution}
Early research in source code authorship attribution focused on the automatic evaluation of students' programming assignments~\cite{leach1995using} and characterizing the authors of programs~\cite{krsul1997authorship}.
Later developments demonstrated a wider range of applications such as safeguarding software integrity and security~\cite{bogomolov2021authorship,alrabaee2017feasibility,alsulami2017source}, malicious code attribution and forensics~\cite{alrabaee2017feasibility}, and identification of legitimate code owners against copyright infringement and plagiarism~\cite{bogomolov2021authorship,prechelt2002finding,shevertalov2009use}. 
Kalgutkar \etal~\cite{kalgutkar2019code} provide a comprehensive review of existing approaches for source code authorship attribution and key challenges in the field.

Several studies have explored the task of identifying code authors using ML/DL techniques.
For instance, Caliskan-Islam \etal \cite{caliskan2015anonymizing} propose an ML-based approach to pinpoint anonymous programmers by studying their coding style, also known as \emph{code stylometry}.
Their solution involves training an ML model on extracted features from the source programs' abstract syntax tree (AST) to capture stylistic patterns in C/C++ code.
By employing methods like random forests, they achieve a remarkable 94\% accuracy on a large Google Code Jam (GCJ) dataset featuring 1,600 programmers and 98\% accuracy on a smaller dataset of 250 programmers, surpassing previous code stylometry research.
Bogomolov \etal~\cite{bogomolov2021authorship} explore path context in AST source code representations and propose an attribution model based on random forests and deep neural networks to characterize and identify authors.
Li \etal \cite{li2023robin} study the interpretability of authorship attribution classifiers through the lens of Siamese neural network models.
Abuhamad \etal \cite{abuhamad2018large} leverage recurrent neural networks to analyze code structure and propose an authorship attribution model, which is effective irrespective of programming language, achieving over 90\% accuracy on large datasets with thousands of programmers.
Li \etal \cite{li2022ropgen} study adversarial training to produce a robust attribution model against malicious misattribution attacks.
These prior ML/DL-based approaches face two common challenges that we seek to circumvent in this work: their reliance on extensive manually labeled datasets for model training and the resulting models' lack of generalization capability against unseen authors and programming languages.

\subsection{Large Language Models in Software Engineering}
LLMs are a type of artificial intelligence (AI) application that can handle complex tasks like recognizing and generating text, source codes, images, and even videos.
This is achieved through training on massive datasets of text and code snippet~\cite{ChatGPT_Doc,team2023gemini}, which empower the models with capabilities for various tasks without task-specific training, \eg creating different creative text formats, writing codes, and answering questions in informative ways~\cite{zhao2023survey,chang2024survey}.

State-of-the-art LLMs, such as ChatGPT and Llama, have shown great potential in programming language processing (PLP) tasks. 
For example, LLMs have been leveraged in code comprehension and generalization~\cite{jain2022jigsaw,liu2023refining,yan2024investigating} and program testing~\cite{liu2023chatting,deng2023large}.
In addition to general-purpose LLMs, there are also models that are specially trained or fine-tuned with domain-specific datasets~\cite{chen2021evaluating,feng2020codebert,tufano2019empirical}.
For example, Codex is an adaptation of GPT-3 tailored for programming tasks~\cite{chen2021evaluating}, which was developed to aid developers by suggesting code snippets, completing code lines, and generating entire functions based on comments or partial code. 
CodeBERT~\cite{feng2020codebert} is a transformer-based model specializing in natural language understanding and generation according to the coding contexts, specifically for tasks like code summarization and documentation. BERT4Bugs~\cite{tufano2019empirical} is another transformer-based model that utilizes BERT architecture to automatically identify and fix bugs in programming code, leveraging large datasets of buggy and corrected code examples.
In this work, we use state-of-the-art general-purpose LLMs for our empirical study, leaving code authorship task-specific fine-tuning out of scope.

\subsection{Authorship Analysis with Large Language Models}
A closely related task to our present study is the use of LLMs for \emph{natural language} authorship analysis.
Huang \etal~\cite{huang2024can} investigated the capability of LLMs in authorship analysis for English texts.
Their motivation comes from the increasing demand for precise text authorship identification, which is essential for tasks such as validating content authenticity (including detecting plagiarism) and combating the dissemination of misinformation.
Unlike conventional techniques which rely heavily on manually engineered stylistic features, Huang \etal demonstrate that existing LLMs are capable of performing authorship analysis tasks without additional training on a domain-specific training corpus.
This suggests that LLMs can analyze stylistic characteristics within text data to distinguish between different authorship styles.

Inspired by the promising results of Huang \etal~\cite{huang2024can}, we investigate LLMs for authorship analysis of \emph{source code}. 
To the best of our knowledge, we are the first to systematically evaluate LLMs for source code authorship attribution.

\section{Empirical Study}\label{sec:empirical}

\begin{figure*}[t]
    \centering
    \includegraphics[width=\textwidth]{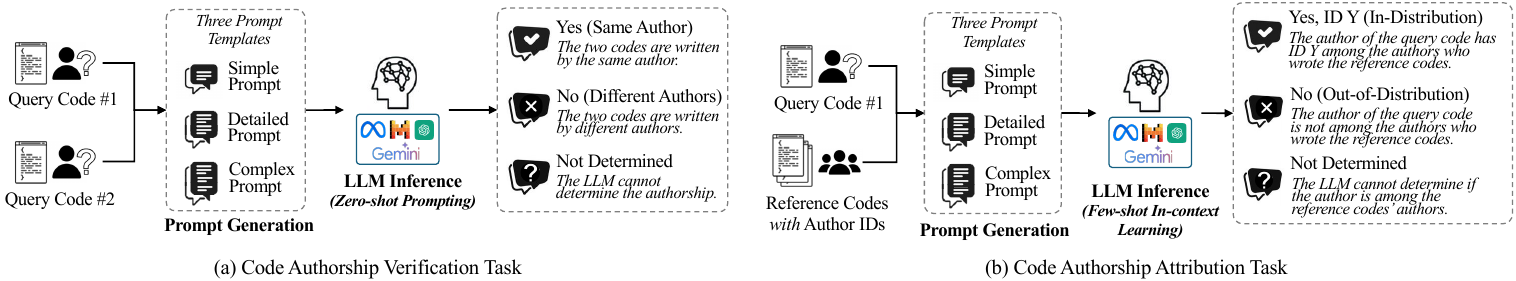}
    \caption{An overview of LLM-based code authorship attribution.}
    \label{fig:overview}
\end{figure*}

Our empirical study seeks to broadly explore the capability of the mainstream LLMs for source code authorship attribution.
An overview of the LLM query pipeline is shown in Fig.~\ref{fig:overview}.
We will start by investigating~\textbf{RQ1} and \textbf{RQ2}.


\subsection{Experiment Setup}

\paragraph{Model Selection and Experiment Environment}
Our empirical study covers four mainstream LLM families, namely OpenAI's GPT~\cite{ChatGPT}, Meta's Llama~\cite{llama}, Mistral~\cite{jiang2023mistral7b}, and Google's Gemini~\cite{gemini}.
The GPT and Gemini models are closed-source, so we conduct experiments using their official APIs.
For the remaining two model families (Llama and Mistral), we use the latest versions as of May 2024 and ran our experiments on a workstation with Ubuntu 22.04 LTS OS, an Intel Xeon Platinum 8368Q CPU, and an NVIDIA RTX A100 80GB GPU.
Overall, we ran experiments on eight LLM models, including GPT 3.5 Turbo, GPT 4o, Llama2 Chat 7B, Llama2 Chat 70B Quantized (GPTQ), Llama3 8B, Llama3 8B Instruct, Mistral2.5 7B Quantized (GPTQ), and Gemini 1.5 Pro.\footnote{Hereafter, GPT-3.5-t, GPT-4o, Llama2-7b, Llama2-70b, Llama3-8b, Llama3-8b-i, Mistral-2.5-7b, and Gemini-1.5-p, respectively.}
Following Huang \etal~\cite{huang2024can}, we set the values of temperature to 0 and top\_p to 1 for all models. All other hyperparameters are set at their default values.

\paragraph{Datasets}
\textcolor{black}{For this study, we experimented with two datasets to answer our RQs.
The first data set was taken from Google Code Jam (GCJ)\footnote{The GCJ is a programming competition where participants solve identical algorithmic challenges, allowing us to examine code written by different authors for the same task.}~\cite{GoogleCodeJam} and the second was a dataset obtained by extracting GitHub repositories that met our criteria (detailed below).
We conducted our small-scale experiments (cf. \textbf{RQ1} and \textbf{RQ2}) using the GCJ 2017 dataset, which is commonly used in research on ML/DL techniques for code authorship attribution~\cite{quiring2019misleading}.
Since C++ represents the largest portion of authors in the GCJ dataset, we focused our evaluation on the C++ subset, which includes 1,632 code samples from 204 authors. 
Later, we also evaluate generalization to other programming languages (cf. \textbf{RQ5}), for which we used the GCJ Java subset containing 2,202 code samples from 74 authors.
For our large-scale experiments (cf. \textbf{RQ3}), we additionally crawled code from public GitHub repositories. 
For the crawling process, we restricted the collection to repositories with a single contributor, containing more than eight C++ code files, ranging from 17 to 300 lines of code, and committed between May and October 2024. This resulted in 26,355 code samples from 500 authors.
Similarly, we crawled Java code from public GitHub repositories in the same period, gathering 55,267 code samples from 686 authors.
An overview of both the GCJ and Github datasets is given in Table~\ref{tab:dataset}.}


\begin{table*}[t]
    \centering
\def\arraystretch{1.1}
\setlength{\tabcolsep}{10pt}
        \caption{\textcolor{black}{Overview of datasets used in experiments, including number of authors, code details, and data collection period.}}    \label{tab:dataset}
            \begin{tabular}{lccccc}      
            \Xhline{2\arrayrulewidth}
            \ccd{\textbf{Dataset}} & \ccd{\textbf{Language}} & \ccd{\textbf{Total \# of Authors}} & \ccd{\textbf{Total \# of Codes} } & \ccd{\textbf{LoC (Min/Max/Ave)}} & \ccd{\textbf{Collection Period}} \\
            \Xhline{2\arrayrulewidth}                                    
            \cc{GCJ C++}    & \cc{C++} & \cc{204} & \cc{1,632}  &\cc{17 / 252 / 67.8} & \cc{2017}                 \\   
            GCJ Java        & Java     & 73       & 2,202       & 4 / 732 / 112.5 & 2017                  \\  
            \cc{GitHub C++} & \cc{C++} & \cc{500} & \cc{26,355} &\cc{17 / 300 / 92.4} & \cc{May--Oct 2024}             \\
            GitHub Java     & Java     & 686      & 55,267      & 17 / 300 / 72 & May--Oct 2024              \\
            \Xhline{2\arrayrulewidth}
            \end{tabular}
\end{table*}


\paragraph{Metrics}
We use the Matthews Correlation Coefficient (MCC)~\cite{chicco2021matthews} as the main evaluation metric.
Its definition takes into account both correct, \ie true positive (TP) and true negative (TN), and incorrect, \ie false positive (FP) and false negative (FN) predictions, as shown below:\footnote{In some cases, the LLM may return an indeterminate answer like ``unsure''. For accuracy and MCC score calculations, we always treat these as wrong answers, \ie either FP or FN.}
\par\nopagebreak\noindent\ignorespaces
{\footnotesize\begin{align*}
\mathrm{MCC}=\frac{(\mathrm{TP} \times \mathrm{TN}-\mathrm{FP} \times \mathrm{FN})}{\sqrt{(\mathrm{TP}+\mathrm{FP}) \times (\mathrm{TP}+\mathrm{FN}) \times (\mathrm{TN}+\mathrm{FP}) \times (\mathrm{TN}+\mathrm{FN})}}
\end{align*}}

The MCC score ranges from -1 to 1, where 1 indicates perfect classification, 0 represents the performance of random guessing, and -1 indicates complete disagreement between the predicted and actual labels.
Compared with metrics like the F1 score, which focuses on precision and recall, the MCC offers a more balanced assessment by considering all entries of the confusion matrix (TP, TN, FP, FN).
This is better suited for evaluating LLMs as they tend to accept user claims in prompts, \ie positive responses, rather than carrying out a critical analysis~\cite{du2024llms}.

\paragraph{Baselines}
As a point of comparison, we employed two existing ML/DL-based code authorship attribution approaches~\cite{caliskan2015anonymizing,abuhamad2018large} and used the artifacts provided by Quiring \etal~\cite{quiring2019misleading} to implement these models.
We record an attribution accuracy of 84.98\% from~\cite{abuhamad2018large} and 90.50\% from~\cite{caliskan2015anonymizing} by testing their models on the GCJ 2017 C++ dataset.
It is worth noting that these values were computed using a leave-one-out protocol to partition the full dataset into train and test sets.
Thus, these numbers are not directly comparable to our accuracy results for LLMs, which use zero- and few-shot prompting.
Nevertheless, we include these numbers to give a sense of current performance for ML/DL models.

\subsection{Prompting Strategies}
\label{sec:Three-Prompts}
The narration and complexity of prompts, so-called \textit{prompt engineering}~\cite{white2023prompt, sahoo2024systematic}, play a vital role in determining LLMs' performance and effectiveness.
We crafted three types of prompts, namely \textit{simple}, \textit{detailed}, and \emph{complex prompts}, to explore how different levels of instructive detail in a prompt can influence outcomes in code authorship tasks.

\paragraph{Simple Prompts}
Simple prompts aim to provide the LLMs with instructions in the most direct and straightforward manner. These prompts are devoid of any additional context or guidance and simply describe the task at hand, \eg the prompt ``Identify the author of the following code snippet from among these candidates.'' may be used for authorship attribution.
After being instructed with a simple prompt, LLMs respond by relying on their pre-trained capabilities and a general understanding of the query, code, and language patterns.

\paragraph{Detailed Prompts}
Moving a step up in prompt sophistication, detailed prompts are designed to include specific features that could be pertinent to identifying code authorship.
These prompts can offer the LLMs more context and background about what aspects of the code might be relevant.
Based on prior work~\cite{abuhamad2018large,caliskan2015anonymizing}, we identified three types of features that can be used for code authorship attribution purposes, namely \emph{layout features}, \emph{lexical features}, and \emph{syntactic features}.
Accordingly, LLMs are instructed to analyze these features as part of our detailed prompts.

\paragraph{Complex Prompts}
Complex prompts go beyond detailed prompts by incorporating an even richer set of features and context.
Here, we leveraged ChatGPT to ask for as many source code characteristics and features as possible, covering a broad range of specific stylistic and structural elements, \eg commenting style, indentation patterns, and the frequency of specific functions or libraries used.
By incorporating these instructions in our prompts, we aim to push the LLMs towards utilizing a more comprehensive array of information and test their ability to integrate and analyze diverse code features to determine authorship accurately.

\subsection{RQ1: Code Authorship Verification with Zero-Shot Prompts}
\label{sec:Zero-Shot}

\begin{figure}[t]
    \centering
    \includegraphics[width=1\linewidth]{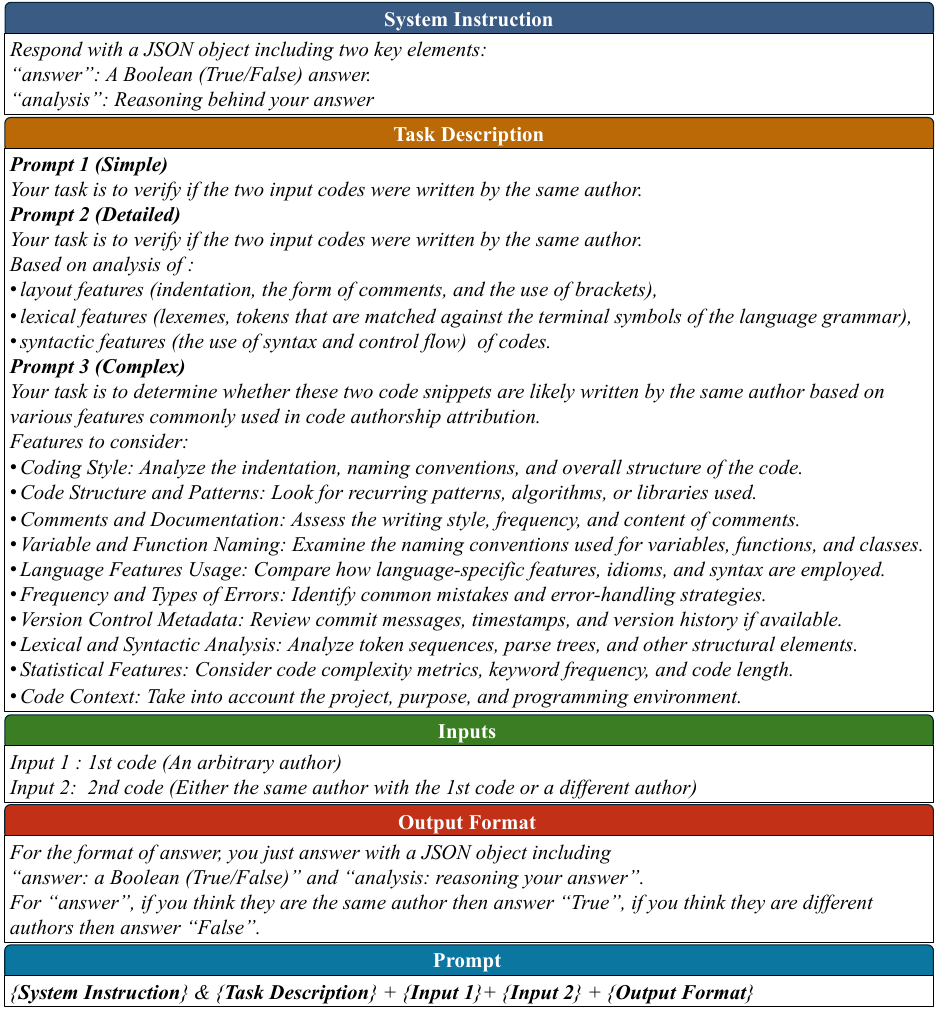}
    \caption{Prompt templates (\textit{Simple}, \textit{Detailed}, and \textit{Complex}) for the zero-shot code authorship verification experiment.}
    \label{fig:verification}
\end{figure}


Our first set of experiments is designed to examine LLMs' capabilities for determining whether two code samples were written by the same author, a task known as code authorship verification.
Here, we adopt a zero-shot prompting approach to assess the eight LLMs; specifically, we randomly sampled a test set consisting of 100 code pairs belonging to the same author for different tasks and another 100 code pairs belonging to different authors (also for different tasks).
Each LLM is given either a test ``same author'' pair or a ``different author'' pair and prompted to answer whether they were written by the same author according to three prompt templates of increasing prompt complexity (P1--P3), as shown in Fig.~\ref{fig:verification}.

For this initial experiment, a detailed confusion matrix of the results is given in Tab.~\ref{tab:zeroshot-cpp-confusion-matrix} to illustrate some of the potential failure cases for our LLM queries; the accuracy and MCC scores are in Tab.~\ref{tab:zeroshot-cpp-accuracy-mcc}.
We present our observations next.

\begin{table}[t]
\def\arraystretch{1.1}
\setlength{\tabcolsep}{2.3pt}
    \centering
        \caption{Confusion matrix for LLM-based code authorship verification with zero-shot prompts over 200 random C++ code sample pairs.
        For prompts where the entries (TP, FN, TN, FP) sum to a value below 200, the LLM returned indeterminate answers for the remaining cases.}
        \label{tab:zeroshot-cpp-confusion-matrix}
            \begin{tabular}{lcccccccccccccc}
            \Xhline{2\arrayrulewidth}
            \ccd{} & \multicolumn{4}{c}{\ccd{\textbf{P1}}} & \ccd{} & \multicolumn{4}{c}{\ccd{\textbf{P2}}} & \ccd{} & \multicolumn{4}{c}{\ccd{\textbf{P3}}}  \\ \hhline{~----~----~----} 
            \ccd{\textbf{Models}}    & \ccd{\textbf{TP}}  & \ccd{\textbf{FN}}  & \ccd{\textbf{TN}}  & \ccd{\textbf{FP}} &\ccd{} & \ccd{\textbf{TP}}  & \ccd{\textbf{FN}}  & \ccd{\textbf{TN}}  & \ccd{\textbf{FP}} & \ccd{} & \ccd{\textbf{TP}}  & \ccd{\textbf{FN}}  & \ccd{\textbf{TN}}  & \ccd{\textbf{FP}} \\ \Xhline{2\arrayrulewidth}
            Llama2-7b           &      98 &       1 &       0 &     100 &      &      99 &       1 &       0 &    100 &       &     {-} &     {-} &     {-} &     {-} \\ 
            \cc{Llama2-70b}     & \cc{14} & \cc{83} & \cc{89} & \cc{7} &\cc{} & \cc{11} & \cc{86} & \cc{94} & \cc{4} & \cc{} & \cc{19} & \cc{81} & \cc{89} & \cc{11} \\
            Llama3-8b           &      12 &      1 &       1 &      16 &      &       8 &      0 &       2 &     9 &       &      10 &      0 &       0 &     7 \\   
            \cc{Llama3-8b-i}    & \cc{70} & \cc{30} & \cc{70} & \cc{30} &\cc{} & \cc{82} & \cc{18} & \cc{59} & \cc{41} & \cc{} & \cc{63} & \cc{37} & \cc{85} & \cc{15} \\
            Mistral-2.5-7b      &      11 &      89 &      87 &      12 &      &      71 &      28 &      42 &     56 &       &      54 &      44 &      50 &      48 \\   
            \cc{Gemini-1.5-p} & \cc{67} & \cc{33} & \cc{96} &  \cc{4} &\cc{} & \cc{88} & \cc{12} & \cc{90} & \cc{10} & \cc{} & \cc{89} & \cc{11} & \cc{84} & \cc{16} \\
            GPT-3.5-t           &       7 &      93 &     100 &       0 &      &       7 &      93 &     100 &      0 &       &       8 &      92 &     100 &       0 \\   
            \cc{GPT-4o}         & \cc{84} & \cc{16} & \cc{92} &  \cc{8} &\cc{} & \cc{79} & \cc{21} & \cc{98} &  \cc{2} & \cc{} & \cc{94} &  \cc{6} & \cc{76} & \cc{24} \\
            \Xhline{2\arrayrulewidth}
            \end{tabular}
\end{table}
~
~
\begin{table}[h]
\def\arraystretch{1.1}
\setlength{\tabcolsep}{2.6pt}
    \centering
        \caption{Accuracy and MCC scores of LLM-based C++ code authorship verification with zero-shot prompts (results of the top two models are shown in \textbf{bold}).}
        \label{tab:zeroshot-cpp-accuracy-mcc}
            \begin{tabular}{lccccccccc}
            \Xhline{2\arrayrulewidth}
            \ccd{} & \multicolumn{4}{c}{\ccd{\textbf{Accuracy(\%)}}} & \ccd{} & \multicolumn{4}{c}{\ccd{\textbf{MCC Scores}}} \\ \hhline{~----~----} 
            \ccd{\textbf{Models}} & \ccd{\textbf{P1}} & \ccd{\textbf{P2}} & \ccd{\textbf{P3}} & \ccd{\textbf{Average}} & \ccd{} & \ccd{\textbf{P1}} & \ccd{\textbf{P2}} & \ccd{\textbf{P3}} & \ccd{\textbf{Average}} \\ \Xhline{2\arrayrulewidth}
            Llama2-7b           & 49.0               & 49.5               & {-}                & 49.3               &       & -0.10              & -0.07              & {-}                & -0.09 \\ 
            \cc{Llama2-70b}     & \cc{51.5}          & \cc{52.5 }         & \cc{54.0}          & \cc{52.7}          & \cc{} & \cc{0.05}          & \cc{0.09}          & \cc{0.11}          & \cc{0.08} \\
            Llama3-8b           & 6.5                & 5.0                & 5.0                & 5.5                &       & -0.88              & -0.90              & -0.90              & -0.89 \\
            \cc{Llama3-8b-i}    & \cc{70.0}          & \cc{70.5}          & \cc{74.0}          & \cc{71.5}          & \cc{} & \cc{0.40}          & \cc{0.42}          & \cc{0.49}          & \cc{0.44} \\
            Mistral-2.5-7b      & 49.0               & 56.5               & 52.0               & 52.5               &       & -0.03              & 0.14               & 0.04               & 0.05 \\
            \cc{Gemini-1.5-p} & \cc{\textbf{81.5}} & \cc{\textbf{89.0}} & \cc{\textbf{86.5}} & \cc{\textbf{85.7}} & \cc{} & \cc{\textbf{0.66}} & \cc{\textbf{0.78}} & \cc{\textbf{0.73}} & \cc{\textbf{0.72}} \\
            GPT-3.5-t           & 53.5               & 53.5               & 54.0               & 53.7               &       & 0.19               & 0.19               & 0.20               & 0.20                        \\
            \cc{GPT-4o}         & \cc{\textbf{88.0}} & \cc{\textbf{88.5}} & \cc{\textbf{85.0}} & \cc{\textbf{87.2}} & \cc{} & \cc{\textbf{0.76}} & \cc{\textbf{0.78}} & \cc{\textbf{0.71}} & \cc{\textbf{0.75}} \\
            \Xhline{2\arrayrulewidth}
            \end{tabular}
\end{table}

\paragraph{Discussion}
The two best-scoring LLMs in terms of accuracy and MCC were GPT-4o and Gemini-1.5-p; they both achieved an MCC of up to 0.78 and raw accuracy of around $89\%$ using the detailed Prompt 2.
It is worth noting the models must be able to distinguish positive and negative pairs to achieve such scores, \ie these LLMs are indeed capable at code authorship verification.
We also see that adding some level of guiding detail in the prompt (P2), as opposed to no detail (P1) or too much detail (P3), worked better for these LLMs.
Another observation for GPT is the remarkable increase in performance going from GPT-3.5-t to GPT-4o, which indicates that future model iterations may become increasingly suitable for deployment on code authorship tasks.

The performance of Mistral and Llama families was less encouraging, with Mistral-2.5-7b, Llama2-7b, and Llama2-70b performing only slightly better than random guessing (achieving around 51\% accuracy) on the verification based on MCC.
This could be due to a lack of task understanding, \eg from the confusion matrix, we see that Llama2-7b almost always returns ``true'' (same author), even for the false pairs.
Thus, at their current stage of development, not all LLMs are suitable for code authorship tasks.
End-users will need to use our experimental design to determine suitable LLMs for their downstream code authorship applications.

The results for the Llama model family offer some additional insights.
Firstly, Llama2-7b failed to handle P3 since the prompt and code snippet lengths exceeded its input token limitation.
This is a problem that will become more pronounced in our later code attribution tasks.
For Llama3-8b, we observed surprisingly poor performance because it returned ``not sure'' indeterminate answers for most of our queries; as a result, the model's accuracy across all prompts was only around 5\%.
It could be possible that more dedicated prompt engineering would force Llama3-8b to only return ``yes'' or ``no'' answers but we did not pursue this further.
Indeed, Llama3-8b-i, the instruction-tuned Llama variant, performed significantly better and yielded the third-best results among the tested LLMs.

\begin{framed}
\noindent \textbf{Summary of Findings in RQ1:}
Our results indicate that not all of the tested LLMs are capable of determining code authorship and they exhibit various failure modes.
Nevertheless, the top-performing LLMs like GPT-4o (87.2\%) and Gemini-1.5-p (85.7\%) indeed demonstrate the ability to verify pairwise code authorship without requiring any task-specific training (\ie by zero-shot prompting).
\end{framed}

\subsection{RQ2: Code Authorship Attribution with Few-Shot, In-Context Learning Prompts}\label{sec:Few-Shot}

\begin{figure}[t]
    \centering
    \includegraphics[width=1\linewidth]{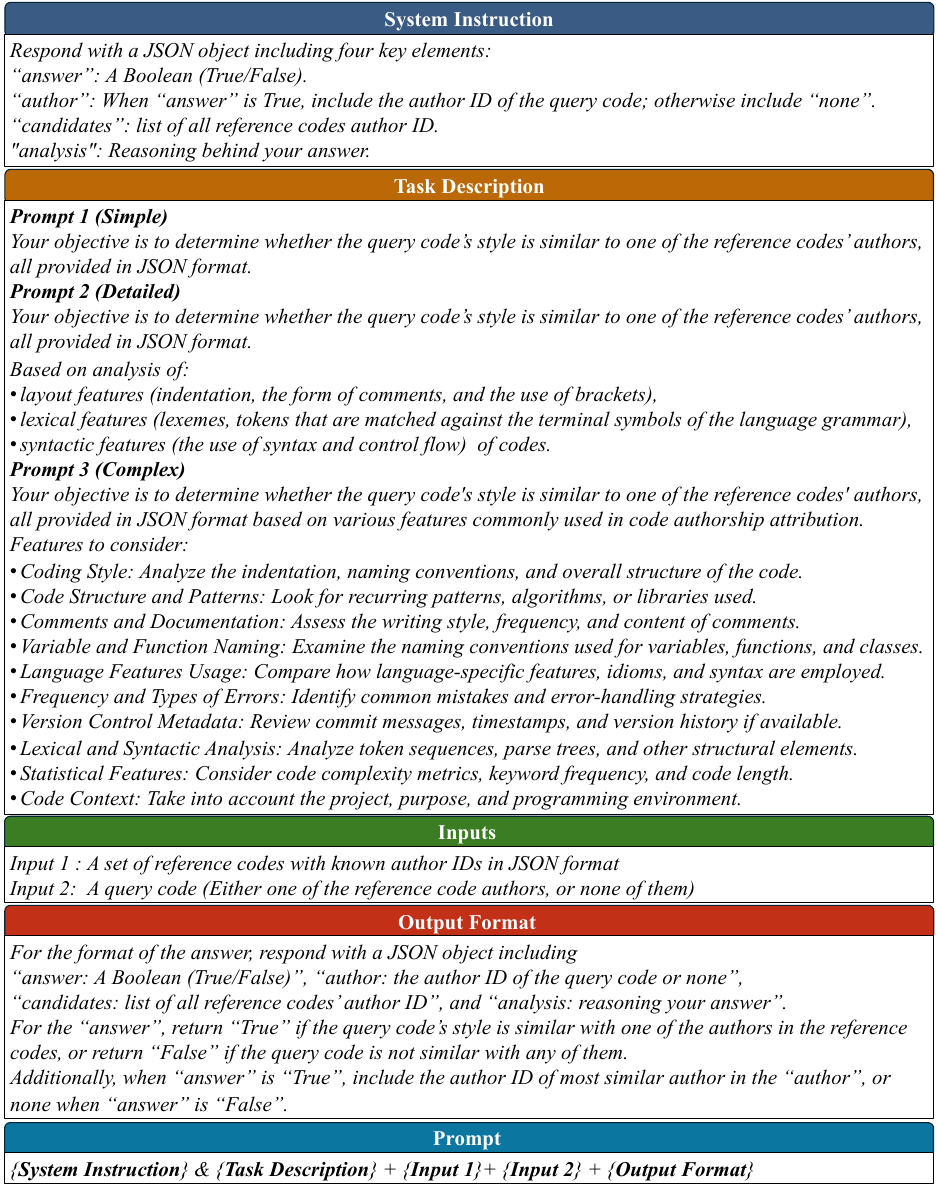}
    \caption{Prompt templates (\textit{Simple}, \textit{Detailed}, and \textit{Complex}) for the few-shot (one-, two-, and three-shot) code authorship attribution experiment.}
    \label{fig:attribution}
\end{figure}

Following the code authorship verification task, we conducted experiments with code authorship attribution using LLMs, focusing attention on Gemini-1.5-p and GPT-4o as they were the most promising models from \textbf{RQ1}.

For code attribution, we supplied a collection of reference code samples for each candidate author to the LLM as part of our prompts; this is followed by a query code sample for which the LLM must attribute to the most likely candidate author from its reference set (or reply that none of the authors match).
The three prompt templates of increasing prompt complexity (P1--P3) are shown in Fig.~\ref{fig:attribution}.

We refer to this setting as ``few-shot'' because we provide $n$ samples per candidate author ($n$ is between one to three) and ``in-context learning'' because the LLM must learn to classify the query from the provided reference samples.
We experimented with the number of candidate authors $k$ set to $3$, $5$, $7$, or $10$.
For each parameter choice of $n$ and $k$, we randomly sampled 100 in-distribution cases, where the queried author is in the candidate set; and 100 out-of-distribution cases, where the queried author is not present in the candidate set.
In either case, the queried author's code sample is from an unseen task.

We present the accuracy and MCC scores in Tab.~\ref{tab:fewshot-cpp-accuracy-mcc}.
For the in-distribution cases, we regard an answer as TP if it correctly identifies the author from the candidate set; conversely, it is FN when the LLM incorrectly returns that the queried author is not among the candidates or when the returned ID is incorrect.
For the out-of-distribution cases, FP and TN are defined as usual.
Due to the input token limitations for LLMs, we were unable to conduct experiments with ten candidate authors in the three-shot setting.
Nevertheless, based on the other results, we may infer that outcomes in this setting will be similar.

\begin{table}
\def\arraystretch{1.1}
\setlength{\tabcolsep}{6pt}
    \centering
        \caption{Accuracy and MCC scores of LLM-based C++ code authorship attribution with one-shot, two-shot, and three-shot prompts over 200 randomly sampled C++ code samples (results with top accuracy and MCC are shown in \textbf{bold}). The value in parentheses indicates number of candidate authors.}
        \label{tab:fewshot-cpp-accuracy-mcc}
            \begin{tabular}{lcccHccccHr}
            \Xhline{2\arrayrulewidth}
            \multicolumn{10}{c}{\ccd{\textbf{One-Shot (one sample per author)}}} & \ccd{} \\  \hhline{-----------}
            \ccd{} & \multicolumn{4}{c}{\ccd{\textbf{Accuracy(\%)}}} & \ccd{} & \multicolumn{4}{c}{\ccd{\textbf{MCC Scores}}} & \ccd{} \\ \hhline{~----~-----} 
            \ccd{\textbf{Models}} & \ccd{\textbf{P1}} & \ccd{\textbf{P2}} & \ccd{\textbf{P3}} & \ccd{\textbf{Average}} & \ccd{} & \ccd{\textbf{P1}} & \ccd{\textbf{P2}} & \ccd{\textbf{P3}} & \ccd{\textbf{Average}} & \ccd{} \\ \Xhline{2\arrayrulewidth}
            Gemini-1.5-p (3)           & 82.0               & \textbf{85.5}      & 82.0               & 83.2               &       & 0.66              & \textbf{0.71}     & 0.66               & 0.68 \\ 
            \cc{Gemini-1.5-p (5)}      & \cc{81.5}          & \cc{78.0}          & \cc{78.0}          & \cc{79.2}          & \cc{} & \cc{0.65}         & \cc{0.56}         & \cc{0.60}          & \cc{0.60} \\
            Gemini-1.5-p (7)           & 78.0               & 79.0               & 78.5               & 78.5               &       & 0.60              & 0.58              & 0.60               & 0.59 \\
            \cc{Gemini-1.5-p (10)}     & \cc{79.5}          & \cc{80.0}          & \cc{78.5}          & \cc{79.3}          & \cc{} & \cc{0.63}         & \cc{0.60}         & \cc{0.61}          & \cc{0.61} \\
            \Xhline{2\arrayrulewidth}
            GPT-4o (3)                   & \textbf{71.0}      & 59.5               & 60.0               & 63.5               &       & \textbf{0.43}     & 0.28              & 0.26               & 0.32 \\
            \cc{GPT-4o (5)}              & \cc{70.5}          & \cc{58.0}          & \cc{59.0}          & \cc{62.5}          & \cc{} & \cc{0.41}         & \cc{0.18}         & \cc{0.20}          & \cc{0.27} \\
            GPT-4o (7)                   & 70.0               & 63.0               & 66.0               & 66.3               &       & 0.41              & 0.27              & 0.33               & 0.34     \\
            \cc{GPT-4o (10)}             & \cc{69.5}          & \cc{64.0}          & \cc{69.0}          & \cc{67.5}          & \cc{} & \cc{0.41}         & \cc{0.28}         & \cc{0.38}          & \cc{0.36} \\
            \Xhline{2\arrayrulewidth}
            \Xhline{2\arrayrulewidth}
            \multicolumn{10}{c}{\ccd{\textbf{Two-Shot (two samples per author)}}} & \ccd{} \\  \hhline{-----------}
            \ccd{} & \multicolumn{4}{c}{\ccd{\textbf{Accuracy(\%)}}} & \ccd{} & \multicolumn{4}{c}{\ccd{\textbf{MCC Scores}}} & \ccd{} \\ \hhline{~----~-----} 
            \ccd{\textbf{Models}} & \ccd{\textbf{P1}} & \ccd{\textbf{P2}} & \ccd{\textbf{P3}} & \ccd{\textbf{Average}} & \ccd{} & \ccd{\textbf{P1}} & \ccd{\textbf{P2}} & \ccd{\textbf{P3}} & \ccd{\textbf{Average}} & \ccd{} \\ \Xhline{2\arrayrulewidth}
            Gemini-1.5-p (3)           & \textbf{88.5}      & 81.0               & 85.5               & 85.0               &       & \textbf{0.77}              & 0.66     & 0.72               & 0.72 \\ 
            \cc{Gemini-1.5-p (5)}      & \cc{85.0}          & \cc{81.0}          & \cc{84.0}          & \cc{83.3}          & \cc{} & \cc{0.70}         & \cc{0.65}         & \cc{0.68}          & \cc{0.68} \\
            Gemini-1.5-p (7)           & 87.5               & 81.5               & 83.0               & 84.0               &       & 0.75              & 0.65              & 0.66               & 0.69 \\
            \cc{Gemini-1.5-p (10)}     & \cc{78.5}          & \cc{76.5}          & \cc{83.5}          & \cc{79.5}          & \cc{} & \cc{0.57}         & \cc{0.54}         & \cc{0.67}          & \cc{0.59} \\
            \Xhline{2\arrayrulewidth}
            GPT-4o (3)                   & 64.5               & 53.5               & 54.0               & 57.3               &       & 0.37              & 0.13              & 0.15               & 0.21 \\
            \cc{GPT-4o (5)}              & \cc{67.0}          & \cc{58.5}          & \cc{61.0}          & \cc{62.2}          & \cc{} & \cc{0.36}         & \cc{0.23}         & \cc{0.32}          & \cc{0.30} \\
            GPT-4o (7)                   & \textbf{71.5}      & 62.0               & 62.0               & 65.2               &       & \textbf{0.44}     & 0.28              & 0.27               & 0.33      \\
            \cc{GPT-4o (10)}             & \cc{63.5}          & \cc{58.0}          & \cc{62.0}          & \cc{61.2}          & \cc{} & \cc{0.27}         & \cc{0.17}         & \cc{0.25}          & \cc{0.23} \\
            \Xhline{2\arrayrulewidth}
            \Xhline{2\arrayrulewidth}
            \multicolumn{10}{c}{\ccd{\textbf{Three-Shot (three samples per author)}}} & \ccd{} \\  \hhline{-----------}
            \ccd{} & \multicolumn{4}{c}{\ccd{\textbf{Accuracy(\%)}}} & \ccd{} & \multicolumn{4}{c}{\ccd{\textbf{MCC Scores}}} & \ccd{} \\ \hhline{~----~-----} 
            \ccd{\textbf{Models}} & \ccd{\textbf{P1}} & \ccd{\textbf{P2}} & \ccd{\textbf{P3}} & \ccd{\textbf{Average}} & \ccd{} & \ccd{\textbf{P1}} & \ccd{\textbf{P2}} & \ccd{\textbf{P3}} & \ccd{\textbf{Average}} & \ccd{} \\ \Xhline{2\arrayrulewidth}
            Gemini-1.5-p (3)           & 84.5                & 77.5               & \textbf{87.0}      & 83.0               &       & 0.71               & 0.60              & \textbf{0.76}      & 0.69 \\ 
            \cc{Gemini-1.5-p (5)}      & \cc{80.5}           & \cc{80.5}          & \cc{84.0}          & \cc{81.7}          & \cc{} & \cc{0.62}          & \cc{0.65}         & \cc{0.69}          & \cc{0.65} \\
            Gemini-1.5-p (7)           & 84.0                & 76.5               & 78.5               & 79.7               &       & 0.69               & 0.57              & 0.58               & 0.61 \\
            \Xhline{2\arrayrulewidth}
            \cc{GPT-4o (3)}              & \cc{\textbf{66.5}}  & \cc{54.5}          & \cc{57.5}          & \cc{59.5}          & \cc{} & \cc{\textbf{0.41}} & \cc{0.18}         & \cc{0.26}          & \cc{0.28} \\
            GPT-4o (5)                   & 64.5                & 61.0               & 60.0               & 61.8               &       & 0.32               & 0.28              & 0.27               & 0.29 \\
            \cc{GPT-4o (7)}              & \cc{63.0}           & \cc{59.5}          & \cc{58.5}          & \cc{60.3}          & \cc{} & \cc{0.27}          & \cc{0.22}         & \cc{0.22}          & \cc{0.24} \\          
            \Xhline{2\arrayrulewidth}
            \end{tabular}
\end{table}

\paragraph{Discussion}
We observe that Gemini-1.5-p always outperforms GPT-4o on both accuracy and MCC for the attribution task across all variations of parameters (number of candidate authors, number of author samples, and prompt complexity).
Across one-, two-, and three-shot settings, Gemini-1.5-p achieved strong accuracy scores of $85.5\%$, $88.5\%$, and $87.0\%$, respectively.
While we cannot directly compare this performance with traditional approaches (see Baseline in Sec.~\ref{sec:empirical}) due to the limited number of few-shot author samples we provided, it remains impressive for the attribution task.
It may be possible that further LLM-specific prompt engineering could improve the scores for GPT-4o---note that GPT-4o performed significantly better than a random guess.

There are also several unexpected observations from these tables.
First, we expected that increasing the number of candidate authors (\eg from $3$ to $10$) would make the tasks increasingly harder for LLMs.
For example, Gemini-1.5-p always had the best results against 3 authors.
However, the remaining experimental results do not always show such a clear trend.
Second, we also expected that increasing the number of code samples per author would make the tasks simpler for LLMs.
This turned out to also not be the case---there were significant variations, again, with no clear trends.
Lastly, whereas the best prompt for GPT-4o was always P1, the best prompt for Gemini-1.5-p ranged from P1 to P3 for different numbers of samples.
Overall, these unexpected variations indicate that a larger test setup and (costly) hyperparameter search may be necessary if one would like to deploy LLMs with optimal sizing parameters for the attribution queries.

\begin{framed}
\noindent \textbf{Summary of Findings in RQ2:}
Both LLMs have shown significant capabilities in code authorship attribution tools using a minimal number of references (\ie few-shot code samples).
Unlike the zero-shot verification setting, Gemini-1.5-p demonstrates better performance across all settings compared to GPT-4o (best accuracy 88.5\% vs. 71.5\%).
On the other hand, we did not observe clear trends in performance when varying the number of reference code samples, selected authors, and choice of prompts.
Deployment of these models may require careful hyperparameter tuning. 
\end{framed}

\section{Scaling Authorship Attribution with Tournament Prompting}\label{sec:ours}
\begin{figure}[t]
    \centering
    \includegraphics[width=0.45\textwidth]{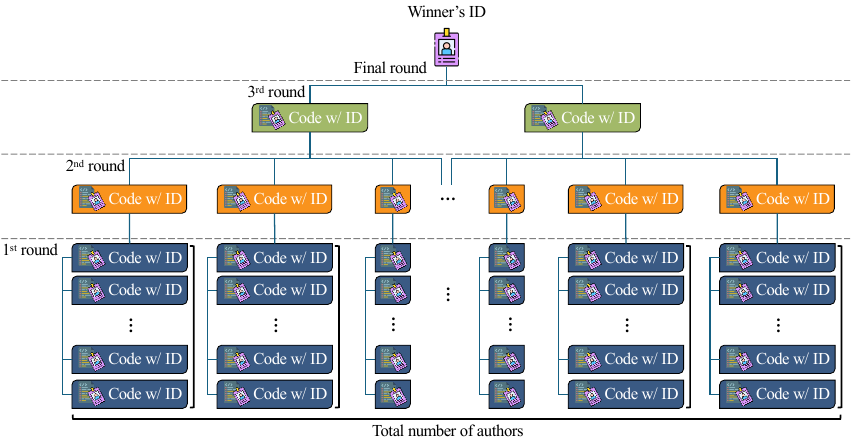}
    \caption{Overview of tournament prompting.}
    \label{fig:tournament_system}
\end{figure}

A common issue we faced in the empirical study (Sec.~\ref{sec:empirical}) was the inherent token limitations of current LLMs in processing lengthy inputs.
This became particularly pronounced when we tried to perform code attribution for a large number of reference author codes simultaneously.
In \textbf{RQ3}, we propose and evaluate a tournament approach that splits the attribution task across many prompts, thus circumventing the input limitation.
An overview of our approach is in Fig.~\ref{fig:tournament_system}.

\subsection{Prompting Methodology}
The tournament authorship attribution process involves the following steps and as shown in Alg.~\ref{alg:tournament}:

\noindentparagraph{Initial Author Pool Selection} Given a large pool of candidate authors $A= \{ a_1, a_2, \ldots, a_n \}$, and a target code snippet $T$, we partition the authors into smaller, evenly-distributed subsets (\eg sample size 12).

\noindentparagraph{Subset Attribution} For each subset, we conduct an authorship attribution query similar to Sec.~\ref{sec:Few-Shot}, where the goal is to determine the likelihood of each candidate being the author of $T$. The comparisons are made by formulating attribution prompts that include the target code snippet and code samples from each candidate author in the subset.

\noindentparagraph{Subset Winner Selection} From each subset, the author with the highest likelihood ($\text{HL}$) is selected to proceed to the next round. This can be formalized as: 
$$\operatorname{winner}_i=\arg \max _{a \in A_i} \text{HL} (T, a)$$
where $A_i$ is the $i$-th subset of authors and $\text{HL}(T, a)$ represents the likelihood given by the LLM for author $a$ being the writer of code snippet $T$.

\noindentparagraph{Iterative Rounds} The winners from each subset form a new pool of candidates. This process is iterated, reducing the pool size with each round.

\noindentparagraph{Final Attribution} In the final round, the remaining authors are compared directly, and the one with the highest $\text{HL}$ score is attributed as the author of the target code snippet.

\begin{algorithm}[t]
\caption{Tournament Prompting Authorship Attribution.}\label{alg:tournament}
\begin{algorithmic}
\small
\Require Target code snippet $T$, Author pool $A$, $sample\_size$
\Ensure Attributed author $Final\_author$
\Function{Get\_HL}{$T, A$}
    \State \textbf{Define a prompt template for LLM}
    \State $prompt \gets$ \Call{Prompt\_Template()}{}
    \State \textbf{Ask LLM for the task with} $T$ \textbf{and} $A$
    \State $Highest\_likelihood\_author \gets $\Call{LLM}{$prompt, T, A$}
    \State \Return $Highest\_likelihood\_author$ 
\EndFunction

\Function{tournament}{$T, A$}
    \State $n\_candidates \gets \Call{len}{A}$
    \State $n\_samples \gets sample\_size$
    \State $advancing\_authors \gets [\,]$
    \For{$start$ \textbf{in} \Call{range}{$0, n\_candidates, n\_samples$}}
        \State $end \gets \Call{min}{start + n\_samples, n\_candidates}$
        \State $author\_list \gets \Call{range}{start, end}$
        \State $Highest\_author \gets$ \Call{Get\_HL}{$T, author\_list$}
        \State $advancing\_authors.\Call{append}{Highest\_author}$
    \EndFor
    \State \Return $advancing\_authors$
\EndFunction

\Function{tournament\_recursive}{$T, A$}
    \State $n\_candidates \gets \Call{len}{A}$
    \If{$n\_candidates \leq 1$}
        \State \textbf{print}("Final Author: ", $A[0]$)
        \State \Return $A[0]$
    \Else
        \State $authors \gets$ \Call{tournament}{$T, A$}
        \State 
        \Return \Call{tournament\_recursive}{$T, authors$}
    \EndIf
\EndFunction

\Function{main}{}
    \State $Final\_author \gets$ \Call{tournament\_recursive}{$T, A$}
    \State \Return $Final\_author$
\EndFunction

\end{algorithmic}
\end{algorithm}

\begin{figure}[t]
    \centering
    \includegraphics[width=0.5\textwidth]{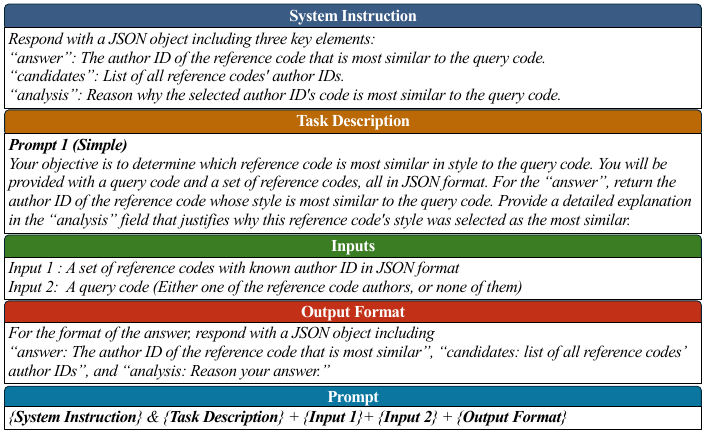}
    \caption{\textcolor{black}{Prompt templates \textit{Simple} for the code authorship experiment with tournament prompting.}}
    \label{fig:tournament}
\end{figure}


In our implementation, we combine ``Subset Attribution'' and ``Subset Winner Selection'' by prompting the LLM to respond directly with the most similar author.

\subsection{RQ3: Large-scale Code Authorship Attribution with Tournament Prompting} \label{sec:tournament}
\textcolor{black}{We evaluated the feasibility of large-scale code authorship using our proposed method against the Github C++ suite of 500 authors.
As before, we used the best performing models, Gemini-1.5-p and GPT-4o, for the tournament experiments.
To facilitate this experiment, we picked a random test set containing 300 random query samples and a corresponding reference set of one-shot reference code samples for each of the 500 authors.
For the tournament, we use author subsets of at most size 12, which fits well into the input token window for the LLMs.
This leads to a total of 4 tournament rounds (500 $\to$ 42 $\to$ 4 $\to$ 1).
Given that simple prompts yielded the most consistent performance in previous experiments, we opted to use only the simple prompt here. The prompt template we used for this experiment is displayed in Fig.~\ref{fig:tournament}.}


Results are presented in Tab.~\ref{tab:Tournament-C++}.
The table shows the accuracy at each round (the second, third, and final round), \ie whether the query sample's author was still present (not eliminated) at that round.
Intuitively, later rounds should be more difficult because the candidate authors that survived to these rounds ought to have higher similarity to the query code.

\begin{table}[t]
    \centering
\def\arraystretch{1.1}
\setlength{\tabcolsep}{10pt}
        \caption{\textcolor{black}{Tournament prompting results for Gemini-1.5-p and GPT-4o were evaluated using GitHub C++ with 300 randomly sampled C++ query code samples.}}    \label{tab:Tournament-C++}
            \begin{tabular}{lcccc}
            \Xhline{2\arrayrulewidth}
            \ccd{\textbf{Models}}               & \multicolumn{3}{c}{\ccd{\textbf{Accuracy(\%)}}} \\ \hhline{~---}
            \ccd{\textbf{Gemini-1.5-p}}       & \ccd{\textbf{2nd round}}              & \ccd{\textbf{3rd round} }               & \ccd{\textbf{Final round}}                     \\
            \Xhline{2\arrayrulewidth}                                    
            \cc{GitHub C++}            & \cc{92}             & \cc{79}             & \cc{65} \\
            \Xhline{2\arrayrulewidth}
            \ccd{\textbf{GPT-4o}}       & \ccd{\textbf{2nd round}}              & \ccd{\textbf{3rd round} }               & \ccd{\textbf{Final round}}                     \\
            \Xhline{2\arrayrulewidth}
            GitHub C++            & 91                &77.3                    & 66                                             \\         
            \Xhline{2\arrayrulewidth}
            \end{tabular}
\end{table}

\paragraph{Discussion}
\textcolor{black}{As before, GPT-4o had slightly weaker performance, except in the final round.
In a sense, the initial round is an ``easy'' authorship task, and our results confirm again that the LLMs have strong capabilities in these smaller-scale code authorship tasks.
However, there is a clear drop in performance for the latter rounds.
This is to be expected because the remaining candidates' perplexity likely increased because the winners of each round are increasingly similar to the query code.
Nevertheless, the final round Top-1 accuracy achieved by both models shows that tournament prompting does indeed work for scaling up code authorship attribution to real-world data.
}

\textcolor{black}{We further observed that, unlike in previous experiments (zero- and few-shot), Gemini-1.5-p and GPT-4o exhibited comparable performance on the real-world GitHub dataset. This may suggest that the models' performance in zero- and few-shot scenarios depends on their training data, the kinds of code sample (algorithmic tasks in GCJ vs. unconstrained code from Github), and that their performance may be affected by potential dataset leakage (for GCJ).}


\begin{framed}
\noindent \textbf{Summary of Findings in RQ3:}
State-of-the-art LLMs (Gemini-1.5-p and GPT-4o) are capable of large-scale, few-shot code authorship attribution with a Top-1 accuracy of \textcolor{black}{around} 65\% using one sample per author when equipped with proper prompting procedures to circumvent their input token length limitations.
\end{framed}

\section{Robustness and Generalization}\label{sec:evaluation}
In this section, we study success criteria for LLM code authorship beyond accuracy metrics, namely \textbf{RQ4} their robustness against adversarial code modifications; and \textbf{RQ5} their generalization capability to different programming languages.

\subsection{RQ4: Robustness against Adversarial Threats}\label{sec:robustness}
To evaluate LLM's robustness against adversarial threats, we tested adversarial code authorship attacks by Quiring \etal~\cite{quiring2019misleading} and Li \etal~\cite{li2022ropgen}. 
The former reported 77.3\% and 81.3\% attack success rates on baseline models~\cite{abuhamad2018large, caliskan2015anonymizing}, while the latter reported 94.8\% success rate against~\cite{abuhamad2018large}. 
\textcolor{black}{We used the GCJ 2017 dataset (C++, 204 authors) to generate modified codes intended to deceive authorship attribution models. This dataset was chosen exclusively for this experiment as the modification results had been validated in prior studies.}
We conducted a zero-shot experiment in two settings: ``Same'' and ``Different'', similar to the previous setup (see Sec.~\ref{sec:Zero-Shot}) to assess the LLMs' resilience to these attacks.

\paragraph{Threat Model}
In the ``same author'' settings, we provided LLMs with two code snippets: a code originally written by the author $A$ for the task $X$, denoted as $T_{A,X}$, and a modified version of $A$'s code for the task $Y$ but styled like a different author $B$, denoted as $f_{B}(T_{A,Y})$.
The transformation $f_{B}(\cdot)$, aims to mislead the attribution model into misattributing the code to $B$, constituting an \emph{evasion} attack.
In the ``different author'' setting, we supplied two codes to the tested LLM: a code originally written by the author $A$ for the task $X$, denoted as $T_{A,X}$, and a modified version of $B$'s code for the task $Y$ written in the style of the author $A$, denoted as $f_{A}(T_{B,Y})$. 
This transformation aims to mimic the coding style of $A$ to mislead the attribution model, constituting an \emph{imitation} attack. 
We used code transformations based on MCTS~\cite{quiring2019misleading} and RoPGen~\cite{li2022ropgen}.

\paragraph{Adversarial-Aware Prompt}
We explored adversarial-aware prompting, where the prompt suggests that the code samples might be altered by evasion or hiding attacks.
On the basis of our proposed prompt templates (see Sec.~\ref{sec:Three-Prompts}), we added a note after the task description indicating that some code samples might be modified: \textit{``Note that some code samples might have been modified using evasion or hiding techniques to alter their stylistic features. Be mindful of these potential modifications and focus on underlying patterns and author-specific traits that remain consistent despite such alterations.''}

Results of na\"ive prompt (same as in \textbf{RQ1}) are in Tab.~\ref{tab:Robustness} and those of the adversarial-aware prompt are in Tab.~\ref{tab:Robustness-adverarial-pmpt}.


\begin{table}[t]
    \centering
    \def\arraystretch{1.1}
        \caption{Robustness experiment with Gemini-1.5-p and GPT-4o over 200 transformed C++ codes via MCTS and RoPGen, respectively; TP \& FN occur as a result of evasion attacks, while TN \& FP arise from imitation attacks.} \label{tab:Robustness}
            \begin{tabular}{lcccc|cc}
            \Xhline{2\arrayrulewidth}
            \ccd{\textbf{Models} }   & \multicolumn{6}{c}{\ccd{\textbf{MCTS}}}  \\  \hhline{~------}
            \ccd{\textbf{Gemini-1.5-p}}  & \ccd{\textbf{TP}}  & \ccd{\textbf{FN}} & \ccd{\textbf{TN}}  & \ccd{\textbf{FP}}  & \ccd{\textbf{Accuracy(\%)}} & \ccd{\textbf{MCC}} \\           
            \Xhline{2\arrayrulewidth}
            P1                         & 17         & 83        & 96        &4          & 56.5          &  0.21    \\
            \cc{P2}                    & \cc{13}    & \cc{87}   & \cc{94}   & \cc{6}    & \cc{53.5}     &  \cc{0.12}     \\
            P3                         & 32         & 68        & 88        & 12        & \bf{60.0}    &  \bf{0.24}  \\
            \Xhline{2\arrayrulewidth}
            \ccd{}                       & \multicolumn{6}{c}{\ccd{\textbf{RoPGen}}} \\   \hhline{~------}
            \ccd{\textbf{Gemini-1.5-p}} & \ccd{\textbf{TP}}  & \ccd{\textbf{FN}} & \ccd{\textbf{TN}}  & \ccd{\textbf{FP}}  & \ccd{\textbf{Accuracy(\%)}} & \ccd{\textbf{MCC}}\\
            \Xhline{2\arrayrulewidth}
            P1                         & 13        & 87         & 97        & 3       &   \bf{55.0}   & \bf{0.18}    \\
            \cc{P2}                    & \cc{12}   & \cc{88}    & \cc{97}   & \cc{3}  &   \cc{54.5}   & \cc{0.17}   \\
            P3                         & 18        & 82         & 92        & 8       &   55.0        & 0.15   \\
            \Xhline{2\arrayrulewidth}
            \Xhline{2\arrayrulewidth}
            \ccd{ }   & \multicolumn{6}{c}{\ccd{\textbf{MCTS}}}  \\  \hhline{~------}
            \ccd{\textbf{GPT-4o}}  & \ccd{\textbf{TP}}  & \ccd{\textbf{FN}} & \ccd{\textbf{TN}}  & \ccd{\textbf{FP}}  & \ccd{\textbf{Accuracy(\%)}} & \ccd{\textbf{MCC}} \\      
            \Xhline{2\arrayrulewidth}
            P1                        & 32         & 68         & 88        & 12      & 60.0        & 0.24   \\
            \cc{P2}                   & \cc{28}    &\cc{72}     & \cc{93}   &\cc{7}   & \cc{60.5}   & \cc{0.28}         \\
            P3                        & 56         & 44         & 75        & 25      & \bf{65.5}   & \bf{0.32}\\
            \Xhline{2\arrayrulewidth}
            \ccd{}                       & \multicolumn{6}{c}{\ccd{\textbf{RoPGen}}} \\   \hhline{~------}
            \ccd{\textbf{GPT-4o}} & \ccd{\textbf{TP}}  & \ccd{\textbf{FN}} & \ccd{\textbf{TN}}  & \ccd{\textbf{FP}}  & \ccd{\textbf{Accuracy(\%)}} & \ccd{\textbf{MCC}}\\

            \Xhline{2\arrayrulewidth}
            P1                        & 31       & 69          & 87         & 13     & \bf{59.0}   & \bf{0.22}   \\
            \cc{P2}                   & \cc{19}  & \cc{81}     & \cc{93}    & \cc{7} & \cc{56.0}   & \cc{0.18}       \\
            P3                        & 44       & 56          & 74         & 26     & 59.0        & 0.19   \\
            \Xhline{2\arrayrulewidth}
            \end{tabular}
\end{table}


\begin{table}[t]
    \centering
\def\arraystretch{1.1}
\setlength{\tabcolsep}{7pt}
        \caption{Robustness experiment using adversarial-aware prompts with Gemini-1.5-p and GPT-4o over 200 transformed C++ codes via MCTS and RoPGen, respectively; TP \& FN occur as a result of evasion attacks, while TN \& FP arise from imitation attacks.} \label{tab:Robustness-adverarial-pmpt}
            \begin{tabular}{lcccc|cc}
            \Xhline{2\arrayrulewidth}
            \ccd{\textbf{Models} }   & \multicolumn{6}{c}{\ccd{\textbf{MCTS}}}  \\  \hhline{~------}
            \ccd{\textbf{Gemini-1.5-p}}  & \ccd{\textbf{TP}}  & \ccd{\textbf{FN}} & \ccd{\textbf{TN}}  & \ccd{\textbf{FP}}  & \ccd{\textbf{Accuracy(\%)}} & \ccd{\textbf{MCC}} \\           
            \Xhline{2\arrayrulewidth}
            P1                         & 20         & 80        & 94        & 6          & 57.0          &  0.21    \\
            \cc{P2}                    & \cc{16}    & \cc{84}   & \cc{94}   & \cc{6}    & \cc{55.0}     &  \cc{0.16}     \\
            P3                         & 27         & 73        & 91        & 9        & \bf{59.0}    &  \bf{0.23}  \\
            \Xhline{2\arrayrulewidth}
            \ccd{}                       & \multicolumn{6}{c}{\ccd{\textbf{RoPGen}}} \\   \hhline{~------}
            \ccd{\textbf{Gemini-1.5-p}} & \ccd{\textbf{TP}}  & \ccd{\textbf{FN}} & \ccd{\textbf{TN}}  & \ccd{\textbf{FP}}  & \ccd{\textbf{Accuracy(\%)}} & \ccd{\textbf{MCC}}\\
            \Xhline{2\arrayrulewidth}
            P1                         & 16        & 84         & 96        & 4       &   \bf{56.0}   & \bf{0.20}    \\
            \cc{P2}                    & \cc{13}   & \cc{87}    & \cc{97}   & \cc{3}  &   \cc{55.0}   & \cc{0.18}   \\
            P3                         & 17        & 83         & 92        & 8       &   54.5        & 0.14   \\
            \Xhline{2\arrayrulewidth}
            \Xhline{2\arrayrulewidth}
            \ccd{ }   & \multicolumn{6}{c}{\ccd{\textbf{MCTS}}}  \\  \hhline{~------}
            \ccd{\textbf{GPT-4o}}  & \ccd{\textbf{TP}}  & \ccd{\textbf{FN}} & \ccd{\textbf{TN}}  & \ccd{\textbf{FP}}  & \ccd{\textbf{Accuracy(\%)}} & \ccd{\textbf{MCC}} \\      
            \Xhline{2\arrayrulewidth}
            P1                        & 46         & 54         & 84        & 16      & 65.0        & 0.32   \\
            \cc{P2}                   & \cc{36}    &\cc{64}     & \cc{90}   &\cc{10}   & \cc{63.0}   & \cc{0.31}         \\
            P3                        & 59         & 41         & 81        & 19      & \bf{70.0}   & \bf{0.41}\\
            \Xhline{2\arrayrulewidth}
            \ccd{}                       & \multicolumn{6}{c}{\ccd{\textbf{RoPGen}}} \\   \hhline{~------}
            \ccd{\textbf{GPT-4o}} & \ccd{\textbf{TP}}  & \ccd{\textbf{FN}} & \ccd{\textbf{TN}}  & \ccd{\textbf{FP}}  & \ccd{\textbf{Accuracy(\%)}} & \ccd{\textbf{MCC}}\\

            \Xhline{2\arrayrulewidth}
            P1                        & 34       & 66          & 82         & 18     & \bf{58.0}   & \bf{0.18}   \\
            \cc{P2}                   & \cc{26}  & \cc{74}     & \cc{87}    & \cc{13} & \cc{56.5}   & \cc{0.16}       \\
            P3                        & 42       & 58          & 73         & 27     & 57.5        & 0.16   \\
            \Xhline{2\arrayrulewidth}
            \end{tabular}
\end{table}

\paragraph{Discussion}
The robustness experiments conducted with Gemini-1.5-p and GPT-4o against transformed C++ codes reveal notable differences in their performance and resilience to adversarial attacks.
Contrary to previous experiments, we found that GPT-4o exhibited greater robustness than Gemini-1.5-p, achieving up to 65.5\% accuracy (\ie, 34.5\% attack success rate) and an MCC of 0.32 for codes modified by MCTS (similarly for RoPGen).
Adversarial-aware prompting enhanced performance against MCTS, especially for GPT-4o, raising accuracy to 70\% and the MCC score to 0.41.
However, the performance slightly dropped in the other cases.
These results suggest that proper adversarial-aware prompting can significantly bolster the robustness of LLMs against adversarial attacks, highlighting the importance of designing sophisticated prompting strategies to counteract adversarial manipulations and improve the reliability of generated outputs.




\begin{framed}
\noindent \textbf{Summary of Findings in RQ4:}
Compared to traditional ML/DL models, LLMs appear to have stronger baseline resilience against adversarial attacks.
Their robustness may be further enhanced (up to 70\% accuracy against MCTS attack) using adversarial-aware prompt strategies.
\end{framed}

\subsection{RQ5: Authorship Attribution with Different Languages}
We selected Java as an alternative programming language to test LLM code authorship attribution capabilities; Java is the second most commonly used programming language in the GCJ dataset.
We repeated our experiments using the CGJ 2017 Java dataset~\cite{GoogleCodeJam} in both the zero-shot and few-shot settings. We also ran the tournament prompting experiments with the GitHub  Java dataset, following the same approach as in the C++ experiments.
Our zero-shot experiments cover the four best-performing models observed from the C++ authorship verification task (\textbf{RQ1}, see Sec.~\ref{sec:Zero-Shot}), the few-shot experiments focus on assessing Gemini-1.5-p and GPT-4o models, as before (\textbf{RQ2}, see Sec.~\ref{sec:Few-Shot}), and the tournament experiments utilize Gemini-1.5-p and GPT-4o models, same as (\textbf{RQ3}, see Sec.~\ref{sec:tournament}).
We only used simple prompt P1 in this set of experiments, as the majority of best performance records were obtained through P1 in previous experiments.
Results of the zero-shot experiments are in Tab.~\ref{tab:zeroshot-java}, few-shot experiments are in Tab.~\ref{tab:few-shot-java}, and tournament experiments are in Tab.~\ref{tab:Tournament-Java}.

\begin{table}[t]
\def\arraystretch{1.1}
\setlength{\tabcolsep}{5pt}
    \centering
        \caption{Confusion matrix, Accuracy, and MCC scores for LLM-based code authorship verification with zero-shot prompt (P1 only) over 200 randomly sampled Java code samples (results of the top two models are shown in \textbf{bold}).}
        \label{tab:zeroshot-java}
            \begin{tabular}{lcccc|cccc}
            \Xhline{2\arrayrulewidth}
            \ccd{\textbf{Models}}    & \ccd{\textbf{TP}}  & \ccd{\textbf{FN}}  & \ccd{\textbf{TN}}  & \ccd{\textbf{FP}} & \ccd{\textbf{Accuracy(\%)}} & \ccd{\textbf{MCC Scores}} \\ \Xhline{2\arrayrulewidth}
            Llama3-8b-i         &      58 &      42 &      96 &      4   & 77.0                  & 0.58      \\   
            \cc{Gemini-1.5-p} & \cc{77} & \cc{23} & \cc{93} &  \cc{7}  & \cc{\textbf{85.0}}    & \cc{\textbf{0.71}}    \\
            GPT-3.5-t           &       9 &      91 &     98  &       2   & 53.5                 & 0.15      \\   
            \cc{GPT-4o}         & \cc{89} & \cc{11} & \cc{95} &  \cc{5}  & \cc{\textbf{92.0}}    & \cc{\textbf{0.84}}    \\
            \Xhline{2\arrayrulewidth}
            \end{tabular}
\end{table}

\begin{table}[t]
\def\arraystretch{1.1}
\setlength{\tabcolsep}{4pt}
    \centering
        \caption{Confusion matrix, Accuracy, and MCC scores for LLM-based code authorship attribution with one-shot, two-shot, and three-shot prompts (P1 only) over 200 randomly sampled Java code samples (results of the top accuracy and MCC are shown in \textbf{bold}). The value in parentheses indicates number of candidate authors.}
        \label{tab:few-shot-java}
            \begin{tabular}{lcccc|cccc}
            \Xhline{2\arrayrulewidth}
            \multicolumn{7}{c}{\ccd{\textbf{One-Shot (one sample per author)}}} \\  \hhline{-------}
            \ccd{\textbf{Models}}    & \ccd{\textbf{TP}}  & \ccd{\textbf{FN}}  & \ccd{\textbf{TN}}  & \ccd{\textbf{FP}} & \ccd{\textbf{Accuracy(\%)}} & \ccd{\textbf{MCC Scores}} \\ 
            \Xhline{2\arrayrulewidth}
            Gemini-1.5-p (3)          &      92 &      8  &      67 &      33   & \textbf{79.5}  & \textbf{0.61}      \\   
            \cc{Gemini-1.5-p (5)}     & \cc{93} & \cc{7}  & \cc{53} &  \cc{47}  & \cc{73.0}      & \cc{0.50}    \\
            Gemini-1.5-p (7)          &      89 &      11 &     56  &       44   & 72.5          & 0.48      \\   
            \cc{Gemini-1.5-p (10)}    & \cc{88} & \cc{12} & \cc{54} &  \cc{46}  & \cc{71.0}      & \cc{0.45}    \\
            \Xhline{2\arrayrulewidth}
            GPT-4o (3)                  &      99 &      1 &      14 &      86   & \textbf{56.5}   & \textbf{0.25}      \\   
            \cc{GPT-4o (5)}             & \cc{98} & \cc{2} & \cc{9}  &  \cc{91}  & \cc{53.5}       & \cc{0.15}    \\
            GPT-4o (7)                  &      99 &      1 &     11  &      89   & 55.0            & 0.21      \\   
            \cc{GPT-4o (10)}            & \cc{98} & \cc{2} & \cc{15} &  \cc{85}  & \cc{56.5}       & \cc{0.23}    \\
            \Xhline{2\arrayrulewidth}
            \Xhline{2\arrayrulewidth}
            \multicolumn{7}{c}{\ccd{\textbf{Two-Shot (two samples per author)}}} \\  \hhline{-------}            
            \ccd{\textbf{Models}}    & \ccd{\textbf{TP}}  & \ccd{\textbf{FN}}  & \ccd{\textbf{TN}}  & \ccd{\textbf{FP}} & \ccd{\textbf{Accuracy(\%)}} & \ccd{\textbf{MCC Scores}} \\ 
            \Xhline{2\arrayrulewidth}
            Gemini-1.5-p (3)          &      97 &      3  &      62 &      38    & \textbf{79.5}       & \textbf{0.63}      \\   
            \cc{Gemini-1.5-p (5)}     & \cc{96} & \cc{4}  & \cc{55} &  \cc{45}   & \cc{75.5}           & \cc{0.56}    \\
            Gemini-1.5-p (7)          &      98 &       2 &     43  &       57   & 70.5                & 0.49      \\   
            \Xhline{2\arrayrulewidth}
            \cc{GPT-4o (3)}             & \cc{100} & \cc{0} & \cc{12}  &  \cc{88}    & \cc{\textbf{56.0}}  & \cc{\textbf{0.25}}    \\
            GPT-4o (5)                  &      100 &      0 &     9    &      91     & 54.5                & 0.22      \\   
            \cc{GPT-4o (7)}             & \cc{100} & \cc{0} & \cc{11}  &  \cc{9}     & \cc{55.5}           & \cc{0.24}    \\
            \Xhline{2\arrayrulewidth}
            \Xhline{2\arrayrulewidth}
            \multicolumn{7}{c}{\ccd{\textbf{Three-Shot (three samples per author)}}} \\  \hhline{-------}                        
            \ccd{\textbf{Models}}    & \ccd{\textbf{TP}}  & \ccd{\textbf{FN}}  & \ccd{\textbf{TN}}  & \ccd{\textbf{FP}} & \ccd{\textbf{Accuracy(\%)}} & \ccd{\textbf{MCC Scores}} \\ 
            \Xhline{2\arrayrulewidth}
            Gemini-1.5-p (3)          &      98  &      2  &      61 &      39   & \textbf{79.5}       & \textbf{0.64}      \\   
            \cc{Gemini-1.5-p (5)}     & \cc{100} & \cc{0}  & \cc{48} &  \cc{52}  & \cc{74.0}          & \cc{0.56}    \\
            \Xhline{2\arrayrulewidth}
            GPT-4o (3)                  &      100 &      0 &      17 &      83   & 58.5                & 0.30      \\   
            \cc{GPT-4o (5)}             & \cc{100} & \cc{0} & \cc{18} &  \cc{82}  & \cc{\textbf{59.0}}  & \cc{\textbf{0.31}}    \\
            \Xhline{2\arrayrulewidth}            
            \end{tabular}
\end{table}

\paragraph{Discussion}
The experiments reveal distinct differences in how each model handles zero-shot and few-shot learning with Java.
GPT-4o excels in zero-shot settings by achieving 92\% accuracy and 0.84 MCC score which is the highest performance of our study, demonstrating strong comprehension and performance without prior examples and modification of prompts. 
However, its performance diminishes significantly in few-shot settings, suggesting difficulties in handling Java language for authorship tasks in few-shot settings.

On the other hand, Gemini-1.5-p shows promising performance across all settings with 85\% accuracy and 0.71 MCC score for zero-shot settings.
It also achieves an accuracy of up to 79.5\% and an MCC score of up to 0.64 for few-shot settings although its effectiveness slightly decreases along with the growth in the number of reference codes.

\textcolor{black}{In the tournament experiment with the GitHub Java dataset, GPT-4o surprisingly delivered the best performance in the tournament setting, achieving 68.7\%. However, Gemini-1.5-p struggled even more with Java code than with C++ samples, attaining only 50\% accuracy.}
These results again suggest the importance of prompt engineering and hyperparameter tuning in the deployment of these LLMs.



\begin{table}[t]
    \centering
\def\arraystretch{1.1}
\setlength{\tabcolsep}{10pt}
        \caption{\textcolor{black}{Tournament prompting results for Gemini-1.5-p and GPT-4o evaluated using GitHub (Java) with 300 randomly sampled Java query code samples.}}    \label{tab:Tournament-Java}
            \begin{tabular}{lcccc}
            \Xhline{2\arrayrulewidth}
            \ccd{\textbf{Models}}               & \multicolumn{3}{c}{\ccd{\textbf{Accuracy(\%)}}} \\ \hhline{~---}
            \ccd{\textbf{Gemini-1.5-p}}       & \ccd{\textbf{2nd round}}              & \ccd{\textbf{3rd round} }               & \ccd{\textbf{Final round}}                     \\
            \Xhline{2\arrayrulewidth}                                    
            \cc{GitHub-Java}            & \cc{87.3}             & \cc{64.3}             & \cc{50.0} \\
            \Xhline{2\arrayrulewidth}
            \ccd{\textbf{GPT-4o}}       & \ccd{\textbf{2nd round}}              & \ccd{\textbf{3rd round} }               & \ccd{\textbf{Final round}}                     \\
            \Xhline{2\arrayrulewidth}
            GitHub-Java            & 93.3                & 80.3                    & 68.7                                             \\         
            \Xhline{2\arrayrulewidth}
            \end{tabular}
\end{table}

\begin{framed}
\noindent \textbf{Summary of Findings in RQ5:}
Gemini-1.5-p shows remarkable performance with the Java programming language across different settings (85\% in zero-shot and 79.5\% in few-shot), while GPT-4o struggles with a lot of false positive cases in few-shot attribution tasks. In contrast, with the real-world GitHub dataset, GPT-4o's performance was better than Gemini-1.5-p.
The overall performance of both models aligns with the experimental outcomes with C++, suggesting that LLMs exhibit comparable and generalizable performance across different programming languages, especially in zero-shot authorship verification scenarios.
\end{framed}

\section{Discussion}\label{sec:discussion}
\subsection{Lessons Learned}




Our experiments reveal several intriguing insights into the behavior and performance of LLMs in code authorship tasks, shedding light on both their capabilities and limitations.

First, we observed that providing more complex guidance in prompts or increasing the number of samples does not always lead to better results (see Sec.~\ref{sec:Few-Shot}). 
For instance, while Gemini-1.5-p performed well with minimal examples in the few-shot settings, its performance declined as the number of references (per author) increased. This suggests that LLMs can suffer from context overload, where the additional information does not enhance and may even hinder the model's ability to accurately attribute code authorship.
A deeper study of the answering consistency (across rounds) and numerical authorship likelihood capabilities of LLMs could give further insights into such differences.

Second, our experiments underscore the importance of prompt engineering in the deployment of LLMs (see Sec.~\ref{sec:empirical}, ~\ref{sec:ours}, and~\ref{sec:evaluation}). Crafting precise and effective prompts can significantly influence the model's performance, especially in context-sensitive tasks (\eg code authorship).
This highlights a critical area for future research and development (\ie optimizing prompt strategies to enhance the performance of LLMs in various settings).

Third, the varying resilience to adversarial attacks between models indicates a need for further advancements in making LLMs more robust in these tasks (see Sec.~\ref{sec:robustness}). While GPT-4o showed greater resilience compared to Gemini-1.5-p, both models exhibited vulnerabilities that could potentially be mitigated through improved prompt design and adversarial prevention (\eg adversarial fine-training).

\textcolor{black}{Fourth, for LLMs such as GPT-4o and Gemini-1.5-p, we utilized their commercial APIs for authorship queries, which incurred additional expenses.
To control such costs in practice, it is important to understand the unit cost of finding a \emph{single} Top-1 result, \ie given one target code by an unknown author, and one reference code for each of the candidate authors in the database, return the author attributed via tournament prompting.
For our tournament experiment on the Github C++ dataset (500 authors), the unit cost was approximately USD 1.58 for GPT-4o and USD 1.60 for Gemini-1.5-p.
Similarly, for the GitHub Java dataset (686 authors), the cost for each result was approximately USD 1.52 for GPT-4o and USD 1.53 for Gemini-1.5-p.}


In summary, our findings show that LLMs can be successful in code authorship attribution tasks.
However, their ultimate utility in real-world deployment will hinge on costs, sophisticated prompt engineering, model selection, hyperparameter choices, and an understanding of the models' sensitivity to context and adversarial conditions.

\subsection{Threats to Validity}
A notable limitation of this work is the potential exposure of the GCJ dataset to the LLMs used in this experiment.
Since GCJ is a public dataset, it is conceivable that these models may have already been seen and trained as code samples.
In that case, although the tested general-purpose LLMs may not train for specific authorship attribution purposes, these models' performance may still be artificially inflated as they can learn some patterns or authors' coding styles during the training.
\textcolor{black}{Thus, we carried our our tournament prompting experiments using freshly crawled Github datasets (see Table~\ref{tab:dataset}).
Based on publicly known information, the code in this dataset does not overlap with LLMs' training data.
Our experiments on this latter dataset shows that LLMs have remarkable performance in large-scale C++ and Java code attribution.}
However, their generalization capabilities to less popular or even new programming languages remains to be seen.

\subsection{Future Work}
For future work, one can consider cross-language authorship attribution, \eg attributing a query Java code against C++ references.
This can provide further insights into LLMs' generalization capabilities.
This line of research can also delve into creating fine-tuned models capable of understanding and linking coding styles across languages, which would be possible with open-source LLMs.
Additionally, enhancing the explainability and interpretability of LLM predictions can build trust and provide deeper insights into their authorship attribution criteria.
This will involve researching methods to make model decisions more transparent and understandable to users, including methods such as attention visualization~\cite{vig2019multiscale}.

\section{Conclusion}\label{sec:Conclusion}
We conducted various experiments to evaluate the capabilities of LLMs in code authorship attribution tasks, including zero-shot, few-shot, and tournament scenarios. Our study shows that some state-of-the-art LLMs have latent capabilities for code authorship attribution without the need for further task-specific fine-tuning.
Moreover, these capabilities are robust against adversarial evasion and hiding attacks and generalize across languages to both C++ and Java.
These insights open new avenues for future research, mining to refine prompt strategies and enhance the robustness of LLMs in diverse and complex scenarios.

\bibliographystyle{IEEEtran}
\bibliography{main}



\end{document}